\documentclass[letter, traditabstract]{aa}  

\usepackage{graphicx}
\usepackage{txfonts}
\usepackage{epstopdf}
\usepackage{xspace} 
\usepackage{natbib}
\bibpunct{(}{)}{;}{a}{}{,}

\newcommand{\spitzer}{\textit{Spitzer}\xspace}

\newcommand{\herschel}{\textit{Herschel}\xspace}
\newcommand{\micron}{$\mu$m\xspace}

\newcommand{\tbol}{T$_{\rm bol}$\xspace}
\newcommand{\tbole}{T$_{\rm bol,ext}$\xspace}

\def\min{{\rm min}}
\def\max{{\rm max}}
\def\ex{{\rm exp}}
\def\obs{{\rm obs}}

\begin{document}
   
\title{Evolution of column density distributions within Orion~A}

\author{A.\ M.\ Stutz  \inst{1} \and
  J.\ Kainulainen  \inst{1}}

\offprints{stutz@mpia.de}

\institute{Max-Planck-Institute for Astronomy, K\"onigstuhl 17, 69117 Heidelberg, Germany \\
  \email{stutz@mpia.de}} 
\date{Received ; accepted }

\abstract{We compare the structure of star--forming molecular clouds
  in different regions of Orion~A to determine how the column density
  probability distribution function ($N$--PDF) varies with
  environmental conditions such as the fraction of young protostars.
  A correlation between the $N$--PDF slope and Class~0 protostar
  fraction has been previously observed in a low-mass star--formation
  region (Perseus) by Sadavoy; here we test if a similar correlation
  is observed in a high--mass star--forming region.  We use \herschel
  PACS and SPIRE cold dust emission observations to derive a column
  density map of Orion~A.  We use the \herschel Orion Protostar Survey
  (HOPS) catalog for accurate identification and classification of the
  Orion~A young stellar object (YSO) content, including the cold and
  relatively short--lived Class~0 protostars (with a $\sim$\,0.14~Myr
  lifetime).  We divide Orion~A into eight independent 0.25 square
  degree (13.5 pc$^2$) regions; in each region we fit the $N$--PDF
  distribution with a power--law, and we measure the fraction of
  Class~0 protostars.  We use a maximum likelihood method to measure
  the $N$--PDF power--law index without binning the column density
  data.  We find that the Class~0 fraction is higher in regions with
  flatter column density distributions.  We test the effects of
  incompleteness, extinction--driven misclassification of Class~0
  sources, resolution, and adopted pixel--scales.  We show that these
  effects cannot account for the observed trend.  Our observations
  demonstrate an association between the slope of the power--law
  $N$--PDF and the Class~0 fractions within Orion~A.  Various
  interpretations are discussed including timescales based on the
  Class~0 protostar fraction assuming a constant star--formation rate.
  The observed relation suggests that the $N$--PDF can be related to
  an ``evolutionary state'' of the gas.  If universal, such a relation
  permits an evaluation of the evolutionary state from the $N$--PDF
  power--law index at much greater distances than those accesible with
  protostar counts.}

\authorrunning{A.\ Stutz et al.}  \titlerunning{Orion PDF and
  environment}

\maketitle

\section{Introduction} \label{sec:intro}

The structure of molecular clouds can be described by the probability
distribution function ($N$--PDF) of their column densities
\citep[e.g.,][]{kai09}.  Observations \citep[e.g.,][]{hill11,hughes13}
and theoretical studies \citep{pad14} suggest that the $N$--PDF
depends on environmental conditions such as turbulence
\citep[e.g.,][]{pad97}, gravity \citep[e.g.,][]{kle00}, magnetic field
strength \citep[e.g.,][]{mol12}, and star--formation
\citep[e.g.,][]{sch13,kai14,abr15}. In particular, an increasing $N$--PDF
slope has been shown to be correlated with an increasing fraction of
Class~0 protostars in individual clumps within Perseus
\citep{sad13}.  

Here we aim to test if a similar correlation is observed in a
different star--forming region.  We present a study of subregions
within the Orion~A molecular cloud that quantifies the link between
the incidence of Class~0 protostars and the $N$--PDF slope.  We bring
together two data sets: column density measurements using
\emph{Herschel} observations, and an accurate protostellar census from
the \herschel Orion Protostar Survey \citep[HOPS; ][Furlan et al., in
  prep]{stutz13,fischer13}. With this analysis, we quantify spatial
variations in the relationship between column density distributions
and fractions of young protostars within Orion~A.

\section{Column density maps and source catalogs}

\subsection{Column density maps}\label{sec:nh}                 

We present the Orion~A column density, N(H), map (Figs.~\ref{fig:nh}
and \ref{fig:ap_nh}) derived from \herschel 160~\micron to 500~\micron
emission maps calibrated against Planck and IRAS data
\citep{bernard10}.  The data were observed as part of the \herschel
Gould Belt program \citep[][]{polychroni13}. The column density and
temperature maps were derived as in \citet{stutz10,stutz13} and
\citet{laun13}. See Appendix~\ref{sec:ap_nh} for more details.

\subsection{Catalog of Class~0, Class~I, and flat spectrum young
  stellar objects}\label{sec:cat}   

\begin{figure}
  \centering
  \scalebox{0.4}{\includegraphics[trim = 20mm 0mm 0mm 0mm, clip]{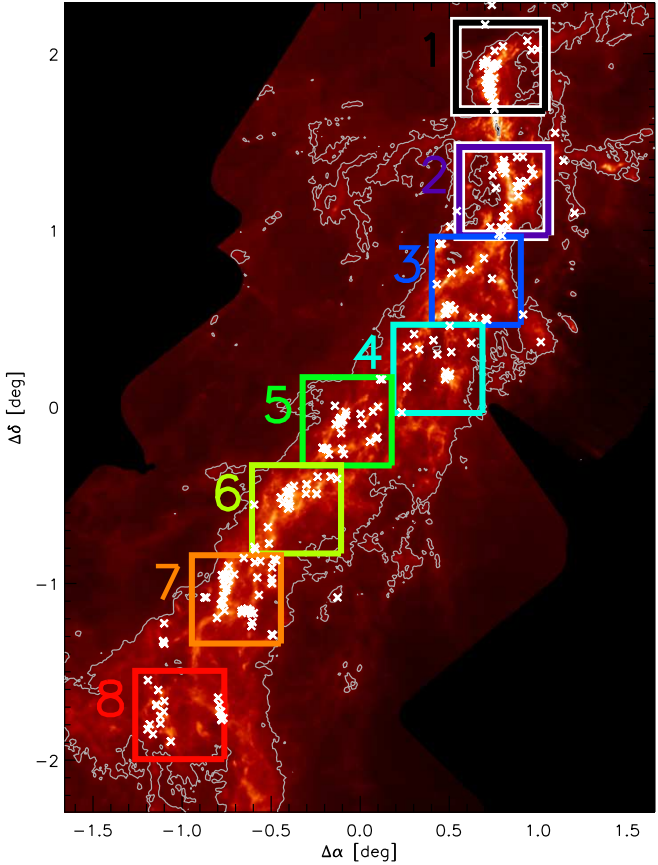}}
  \caption{Orion~A N(H) map shown on a log scale; the
    N(H)$=3.7 \ times 10^{21}$~cm$^{-2}$ contour is indicated in
    grey. The locations of protostars are indicated as
    $\times$--symbols.  Numbered boxes (with 1/4 square degree area or
    3.67~pc on a side) indicate the regions into which we divide the
    Orion~A cloud.}
  \label{fig:nh}
\end{figure}

The young stellar object (YSO) catalog is the union of the PACS Bright
Red Sources (PBRS) sample \citep{stutz13} of extremely young
\herschel--detected Class~0 protostars \citep{heiderman15,vankempen09}
and the HOPS sample \citep[Furlan et al., in prep., and][]{fischer13}
of Class~0, Class~I, and flat spectrum YSOs. The HOPS YSO catalog is
based on the \spitzer catalog from \citet{megeath12}, which excludes
extragalactic and stellar source contamination by means of infrared
colors.  The YSOs have well--sampled SEDs from near--infrared to the
submillimeter wavelengths, including \herschel 70~\micron,
100~\micron, and 160~\micron measurements as well as APEX LABOCA
870~\micron data.  We also include APEX SABOCA 350~\micron data when
available \citep{stanke10,stutz13,safron15} to sample the peak of the
cold envelope emission. The locations of the YSOs are indicated in
Fig.~\ref{fig:nh}.

\begin{figure}
  \centering
  \scalebox{0.71}{\includegraphics[trim = 20mm 0mm 0mm 4mm, clip]{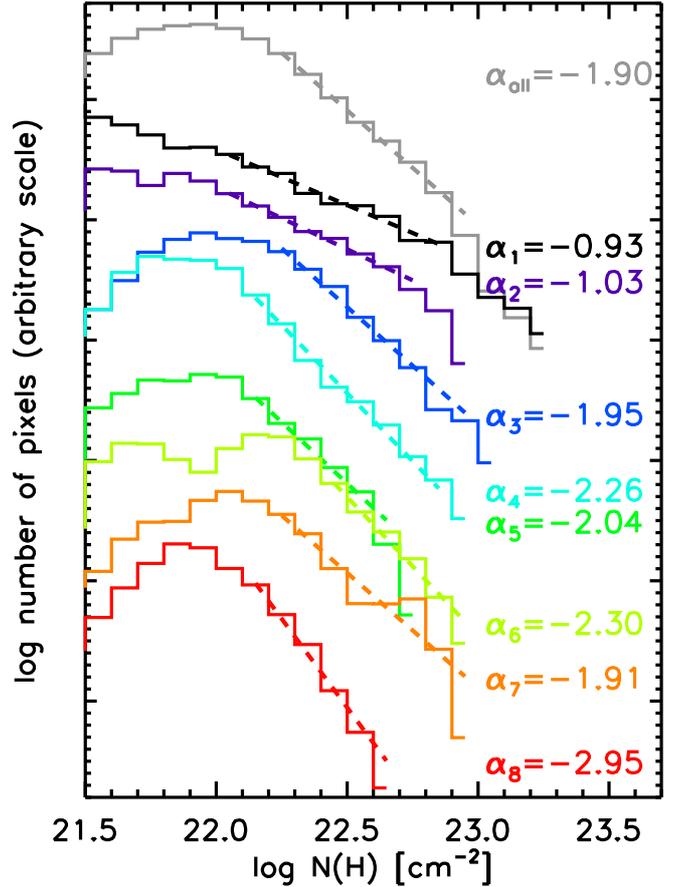}}
  \caption{N(H) distributions for each region shown in
    Fig.~\ref{fig:nh}. The combined N(H) distribution for all
    regions is shown in grey. The dashed curves show the best--fit
    slope assuming a power--law distribution of N(H); each value for
    the slope is listed (see Table~\ref{tab:sf} for slope
    errors). Here we show PDF data of bins containing more than 10
    pixels, and each major tick mark represents a factor of 10.
    Mean-normalization, N(H)/$<$N(H)$>$ in each region, has no effect
    on the derived PDF slopes because the operation simply results in
    a linear translation of the x--axis.}
  \label{fig:pdf}
\end{figure}

\section{Results}

\subsection{Power--law N(H) distributions}\label{sec:fits}  

We analyze the high column density portion of the $N$--PDFs within the
filamentary sub--regions of Orion~A. \citet{lombardi15} demonstrate
that the high--N(H) portion of the $N$--PDF can be reliably inferred
from \herschel observations above $A_{\rm K} \sim 0.2$ mag even when
accounting for complicating factors such as line--of--sight blending
and errors in the background emission or extinction estimates.  We
divide Orion~A into 8 separate square regions that are 3.67~pc on a
side. The region locations are chosen such that they are independent
(have no overlap), avoid the center of the Orion Nebula Cluster (ONC)
which is affected by gross protostellar incompleteness and saturation
in both \spitzer and \herschel observations), and cover regions that
are presently forming stars.  The region locations are shown in
Fig.~\ref{fig:nh}.

\begin{figure}
  \centering
  \scalebox{0.6}{\includegraphics[trim = 20mm 0mm 0mm 4mm, clip]{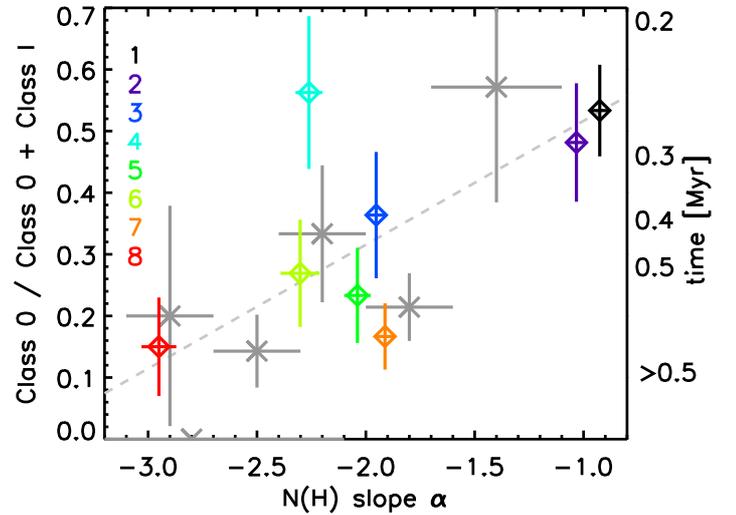}}
  \caption{Correlation between the power--law index of the $N$--PDF
    and Class~0 fraction across Orion~A regions. The Class~0 fraction
    can be related to a timescale as indicated assuming a constant SFR
    (see text).  Grey $\times$--symbols are the data from Perseus
    from~\citet{sad13}, \citet{sadavoy14}, and Sadavoy (private
    communication, 2015) .  Dashed grey curve shows the linear fit to
    the data, which have a correlation coefficient of 0.7.}
  \label{fig:cor}
\end{figure}

We extract the $N$--PDF and fit the high column density portion with a
power--law in each region.  Our results are presented in
Table~\ref{tab:sf}, and Figs.~\ref{fig:pdf} and \ref{fig:cor}.  The
\herschel 250~\micron beam is $\sim\,$18\arcsec\ FWHM; therefore we
use an 18\arcsec pixel scale for extracting the N(H) distributions to
minimize pixel--to--pixel correlations.  Inspection of the $N$--PDFs
reveals that they have an approximate power--law shape above log
N(H)$\,\sim\,$ 22.0 to 22.4 and below log N(H)$\,\sim\,$22.7 to 23.0.
Thus, for each region the definition of the fitting regime is driven by
the requirement of excluding areas with curvature in the
$N$--PDFs. The minimum and maximum N(H) values used to derive the
power--law indices are listed in Table~\ref{tab:sf}.

We fit a power--law to the $N$--PDF using a maximum likelihood method
that does not require binning the data and is therefore free of any
potential errors due to bin size.  We obtain power--law indexes that
vary from $\alpha = -0.93$ to $\alpha = -2.95$, or a factor of about
3. We test the effects of varying the beam size and pixel size on our
derived indices and find that our fitting method is robust.  We also
test the effect of varying the upper bound of the fit and find that
the effect on the indices is much smaller than the measured variations
across regions.  See appendix~\ref{sec:ap_fit} for more details.

\begin{table*}
\caption{Best--fit $N$--PDF power--law slopes and Class~0 protostar fractions}
\begin{center}
\begin{tabular}{cccccccc}
  \hline \hline
  Region$^a$ &
  R.A.\ &
  Decl.\ &
  [Min,Max] log N(H)$^b$ &
  PDF index$^c$ &
  N$_{\rm Class 0}$ &
  N$_{\rm YSO}$ & 
  Fraction$^d$ \\
  \hline
  1 & 05:35:11.91 & -05:01:08.00 & [22.0, 22.9] & -0.93 $\pm$ 0.04 & 24 & 45 & 0.53 $\pm$ 0.07  \\
  2 & 05:35:07.37 & -05:43:23.15 & [22.0, 22.8] & -1.03 $\pm$ 0.04 & 13 & 27 & 0.48 $\pm$ 0.10  \\
  3 & 05:35:44.43 & -06:13:49.45 & [22.2, 23.0] & -1.95 $\pm$ 0.05 & 8 & 22 & 0.36 $\pm$ 0.10  \\
  4 & 05:36:36.13 & -06:43:45.72 & [22.1, 22.9] & -2.26 $\pm$ 0.06 & 9 & 16 & 0.56 $\pm$ 0.12  \\
  5 & 05:38:38.41 & -07:01:35.38 & [22.1, 22.7] & -2.04 $\pm$ 0.06 & 7 & 30 & 0.23 $\pm$ 0.08  \\
  6 & 05:39:46.66 & -07:31:29.36 & [22.4, 23.0] & -2.30 $\pm$ 0.09 & 7 & 26 & 0.27 $\pm$ 0.09  \\
  7 & 05:41:07.68 & -08:02:05.33 & [22.2, 23.0] & -1.91 $\pm$ 0.05 & 8 & 48 & 0.17 $\pm$ 0.05  \\
  8 & 05:42:23.95 & -08:41:22.42 & [22.1, 22.7] & -2.95 $\pm$ 0.08 & 3 & 20 & 0.15 $\pm$ 0.08  \\
  \hline
  All$^e$ &    &      & [22.2, 23.0]  & -1.90 $\pm$ 0.02 & 79 & 234 & 0.34 $\pm$ 0.03 \\
  \hline
\end{tabular}
\end{center}
\footnotesize{$^a$ Region numbers as in Fig.~\ref{fig:nh}; each
  region is 0.5~deg in a side.\\ $^{b}$ Minimum and maximum log N(H)
  values respectively included in the fit to the power--law index of
  the PDF.  \\$^c$ PDF power-law index errors are estimated using the
  $\Delta\,\chi^2 = 1.0$ interval. \\$^d$ Fraction of Class~0
  protostars relative to the total number YSOs: N$_{\rm Class
    0}$/N$_{\rm YSO}$. The fractional errors are derived assuming a
  binomial distribution (see text). \\$^e$ Integrated properties of the
  above regions.}
\label{tab:sf}
\end{table*}

\subsection{Class~0 protostar fractions}\label{sec:frac}

Using the catalogs discussed in \S~\ref{sec:cat}, we count the total
number of YSOs and the subset of Class~0 protostars in each
region (see Table~\ref{tab:sf} and Fig.~\ref{fig:nh}).  We define the
Class~0 fraction as the number Class~0 protostars divided by the total
number of YSOs: N$_{\rm Class0}$ / N$_{\rm YSO}$.  We define
the Class~0 protostars as the subset of sources with bolometric
temperatures \tbol $< 70$~K \citep{chen95}. \tbol is defined as the
temperature of a black--body with the same flux--weighted mean
frequency as the observed SED \citep{myers93}.  The original
protostellar \tbol classification accounted for the effects of
foreground extinction \citep[][; see also Dunham et al.\
2013]{chen95}\nocite{dunham13}.  Here we calculate \tbol based on the
observed SED, without additional extinction corrections.  See text
below and Appendix~\ref{sec:ap_comp} for analysis on the effects of
foreground extinction.  Our wavelength coverage samples the peak of
even the coldest Class~0 protostellar SEDs and allows for a robust
protostellar classification based on \tbol \citep{stutz13,dunham14}.

We estimate the errors on the fraction of Class~0 protostars to the
total number of YSOs using the binomial distribution as follows.
In each region the number of Class~I protostars is (obviously)
N$_{\rm ClassI} = $ N$_{\rm YSO} - $ N$_{\rm Class0}$, where we
include the flat--spectrum YSOs in the Class~I sample.  Since the
YSOs in each region have a probability p of being Class~I and a
probability q = (1 - p) of being Class~0, the expected number of
Class~0 protostars is therefore N$_{\rm YSO} \times {\rm q}$. The
N$_{\rm Class0}$ error is $\sqrt{N_{\rm YSO} \times p \times q}$ and
the fractional error is therefore
$\sigma = \sqrt{p \times q / N_{\rm YSO}}$.  The final numbers of
YSOs, Class~0 protostars, fractions and respective errors in
each region are presented in Table~\ref{tab:sf}.

Two principal effects could potentially alter the measured fractions
of Class~0 protostars: incompleteness and misclassification.
Variations in flux completeness across Orion~A are dominated by the
spatially--varying level of nebulosity. We determine the completeness
limits across the cloud by injecting fake sources into the
PACS~70~\micron images and measuring the flux at which 90\% of the
sources are recovered.  Region 1 has the highest 70~\micron flux
completeness limit.  We apply this limit to the entire YSO 
sample in each region and find that the Class~0 fractions are largely
unaffected by incompleteness. We therefore do not correct the
observational numbers presented in Table~\ref{tab:sf}.  See
Appendix~\ref{sec:ap_comp} and Fig.~\ref{fig:ap_fraccomp} for more
discussion.

With a reliable sample of YSOs and protostars, the main classification
ambiguities that may hinder the identification of Class~0 protostars
include any extrinsic effects (that is, effects acting independently
of the Class~0 envelope) that will cause the observed SED to appear
redder.  The two main cuprits are the inclination of the disk relative
to the line--of--sight (LOS) and foreground reddening.  Using a grid
of protostellar SED models \citep[Furlan et al., in prep.,
][]{stutz13}, we test the effects of both on the \tbol--based
classification. Even assuming the most elevated levels of extinction
measured from our N(H) maps toward the positions of the protostars, we
find that neither effect can account for the variations in the Class~0
fractions reported in \S~\ref{sec:con}.  See
Appendix~\ref{sec:ap_comp} and Figs.~\ref{fig:ap_tbol} and
\ref{fig:ap_cfrac} for more details.

\subsection{The connection between the $N$--PDF and the Class~0 fraction}\label{sec:con}

The observed $N$--PDF slopes and Class~0 fractions show a spatial
variation within Orion~A.  From North to South, the $N$--PDF slope
steepens and the Class~0 fraction decreases.  Fig.~\ref{fig:cor},
which also includes data from \citet{sad13}, \citet{sadavoy14}, and
Sadavoy (private communication, 2015), shows this observed
correlation.  A linear fit to the data results in a slope of 0.2 and
an intercept of 0.7.  We measure a correlation coefficient of 0.7.
According to the t--test there is a 0.3~\% chance of having a
correlation coefficient this high in an uncorrelated random sample.

\section{Discussion}\label{sec:conc}

The observed $N$--PDF slopes and Class~0 fractions indicate clear
spatial variations within a single molecular cloud (Orion~A).  Making
the simple assumption of a constant star--formation rate (SFR), the
Class~0 fraction can be related to time (t).  If the Class~0 and
Class~I lifetimes are 0.1~Myr and 0.4~Myr, respectively, then at
$t < 0.1$~Myr the fraction will be equal to 1; at 0.1~Myr $> t > $
0.5~Myr the fraction is $= [t / (0.1 \rm{Myr})]^{-1}$; at
$t > 0.5$~Myr, the fraction will be constant and equal to
$1 / (1 + t_1/t_0)\,\sim$\,0.26, where $t_0$ and $t_1$ are the
lifetimes of the Class~0 and Class~I phases, respectively
\citep{dunham14}.  Under these assumptions, high fractions correspond
to short evolutionary timescales of the on--going star--formation
event, and the $N$--PDFs change on $\sim\,$0.3~Myr timescales.

However, variations in the SFR could explain the observed trend.  We
know that Orion~A has formed stars for longer than 0.5~Myr
\citep[e.g.,][]{megeath12}. An elevated Class~0 fraction may therefore
reflect an increasing SFR over time compared to regions with lower
Class~0 protostar fractions.

O--star feedback could compress the gas and cause a flattening of the
$N$--PDF slopes.  Regions 1 and 2 would be most affected by feedback
because of the O--star population in the northern portion of Orion~A.
In Perseus O--stars themselves are not responsible because the region
does not contain any such stars.  However, \citet{sadavoy14} propose
that feedback from the low--mass protostars themselves may account for
the observed correlation between Class~0 fraction and $N$--PDF slope.
This mechanism could potentially operate in Orion as well.

\section{Conclusions}

We analyze the $N$--PDF slopes and Class~0 protostellar fractions
within sub--regions of the Orion~A molecular cloud.  Our conclusions
are as follows.

\noindent$\bullet$ We observe a progression from shallow to steep
$N$--PDF slopes from North to South in the Orion~A cloud.  This
progression shows that there is no unique $N$--PDF, but that
the $N$--PDF shape depends on environment.

\noindent$\bullet$ We observe a correlation between increasing
$N$--PDF slope and increasing fraction of Class~0 protostars in
subregions of Orion~A.  Under the assumption that the Class~0 fraction
is related to time via an assumption of a constant SFR, evolutionary
timescales for each region can be derived.  This suggests that regions
with shallower slopes have younger ``evolutionary states''.

If universal, this relation permits an evaluation of the evolutionary
state from the N(H) power--law index measurement, which is possible at
much greater distances than regions that are accesible with protostar
counts.  A key aspect of this study is that the slopes do not change
significantly with resolution; therefore for fixed angular resolution
we expect to obtain the same slope measurement over a broad range of
distances.

\begin{acknowledgements}
  The authors thank Andrew Gould for extremely helpful discussions.
  The authors thank Jean--Philippe Bernard for providing the offsets
  for the \herschel maps.  We thank Neal Evans for a helpful referee
  report.  We are very grateful to the HOPS team for data use in
  advance of publication.  The authors thank H.\ Linz, M.\ Nielbock,
  and A.\ Schmiedeke for assistance with the \herschel data reduction.
  The authors thank E.\ Keto, M.\ Fouesneau, L.\ Hartmann, L.\
  Konstandin, S.\ T.\ Megeath, M.\ Ness, P.\ Myers, T.\ Robitaille and
  S.\ Sadavoy for helpful discussions. The work of AS and JK was
  partially supported by the Deutsche Forschungsgemeinschaft priority
  program 1573 ("Physics of the Interstellar Medium").  We include
  data from \herschel, a European Space Agency space observatory with
  science instruments provided by European-led consortia and with
  important participation from NASA. We use data from the Spitzer
  Space Telescope and the Infrared Processing and Analysis Center
  Infrared Science Archive, which are operated by JPL/Caltech under a
  contract with NASA. We also include data from APEX, a collaboration
  between the Max-Planck-Institut f\"ur Radioastronomie, the European
  Southern Observatory, and the Onsala Space Observatory.
\end{acknowledgements}


\begin{thebibliography}{37}
\expandafter\ifx\csname natexlab\endcsname\relax\def\natexlab#1{#1}\fi

\bibitem[{Abreu {et~al.}(2015)}]{abr15}
Abreu, V. {et~al.} 2015, Astronomy and Astrophysics, 000, 000

\bibitem[{{Aniano} {et~al.}(2011){Aniano}, {Draine}, {Gordon}, \&
  {Sandstrom}}]{aniano11}
{Aniano}, G., {Draine}, B.~T., {Gordon}, K.~D., \& {Sandstrom}, K. 2011, \pasp,
  123, 1218

\bibitem[{{Bernard} {et~al.}(2010){Bernard}, {Paradis}, {Marshall}, {Montier},
  {Lagache}, {Paladini}, {Veneziani}, {Brunt}, {Mottram}, {Martin},
  {Ristorcelli}, {Noriega-Crespo}, {Compi{\`e}gne}, {Flagey}, {Anderson},
  {Popescu}, {Tuffs}, {Reach}, {White}, {Benedettini}, {Calzoletti},
  {Digiorgio}, {Faustini}, {Juvela}, {Joblin}, {Joncas}, {Mivilles-Deschenes},
  {Olmi}, {Traficante}, {Piacentini}, {Zavagno}, \& {Molinari}}]{bernard10}
{Bernard}, J.-P., {Paradis}, D., {Marshall}, D.~J., {et~al.} 2010, \aap, 518,
  L88

\bibitem[{{Bohlin} {et~al.}(1978){Bohlin}, {Savage}, \& {Drake}}]{bohlin78}
{Bohlin}, R.~C., {Savage}, B.~D., \& {Drake}, J.~F. 1978, \apj, 224, 132

\bibitem[{{Chen} {et~al.}(1995){Chen}, {Myers}, {Ladd}, \& {Wood}}]{chen95}
{Chen}, H., {Myers}, P.~C., {Ladd}, E.~F., \& {Wood}, D.~O.~S. 1995, \apj, 445,
  377

\bibitem[{{Dunham} {et~al.}(2013){Dunham}, {Arce}, {Allen}, {Evans},
  {Broekhoven-Fiene}, {Chapman}, {Cieza}, {Gutermuth}, {Harvey}, {Hatchell},
  {Huard}, {Kirk}, {Matthews}, {Mer{\'{\i}}n}, {Miller}, {Peterson}, \&
  {Spezzi}}]{dunham13}
{Dunham}, M.~M., {Arce}, H.~G., {Allen}, L.~E., {et~al.} 2013, \aj, 145, 94

\bibitem[{{Dunham} {et~al.}(2014){Dunham}, {Stutz}, {Allen}, {Evans},
  {Fischer}, {Megeath}, {Myers}, {Offner}, {Poteet}, {Tobin}, \&
  {Vorobyov}}]{dunham14}
{Dunham}, M.~M., {Stutz}, A.~M., {Allen}, L.~E., {et~al.} 2014, ArXiv e-prints:
  1401.1809

\bibitem[{{Fischer} {et~al.}(2013){Fischer}, {Megeath}, {Stutz}, {Tobin},
  {Ali}, {Stanke}, {Osorio}, {Furlan}, {HOPS Team}, \& {Orion Protostar
  Survey}}]{fischer13}
{Fischer}, W.~J., {Megeath}, S.~T., {Stutz}, A.~M., {et~al.} 2013,
  Astronomische Nachrichten, 334, 53

\bibitem[{{Gould}(1995)}]{gould95}
{Gould}, A. 1995, \apj, 440, 510

\bibitem[{{Heiderman} \& {Evans}(2015)}]{heiderman15}
{Heiderman}, A. \& {Evans}, II, N.~J. 2015, ArXiv e-prints

\bibitem[{{Hill} {et~al.}(2011){Hill}, {Motte}, {Didelon}, {Bontemps},
  {Minier}, {Hennemann}, {Schneider}, {Andr{\'e}}, {Men'shchikov}, {Anderson},
  {Arzoumanian}, {Bernard}, {di Francesco}, {Elia}, {Giannini}, {Griffin},
  {K{\"o}nyves}, {Kirk}, {Marston}, {Martin}, {Molinari}, {Nguyen Luong},
  {Peretto}, {Pezzuto}, {Roussel}, {Sauvage}, {Sousbie}, {Testi},
  {Ward-Thompson}, {White}, {Wilson}, \& {Zavagno}}]{hill11}
{Hill}, T., {Motte}, F., {Didelon}, P., {et~al.} 2011, \aap, 533, A94

\bibitem[{{Hughes} {et~al.}(2013){Hughes}, {Meidt}, {Schinnerer}, {Colombo},
  {Pety}, {Leroy}, {Dobbs}, {Garc{\'{\i}}a-Burillo}, {Thompson}, {Dumas},
  {Schuster}, \& {Kramer}}]{hughes13}
{Hughes}, A., {Meidt}, S.~E., {Schinnerer}, E., {et~al.} 2013, \apj, 779, 44

\bibitem[{Kainulainen {et~al.}(2009)Kainulainen, Beuther, Henning, \&
  Plume}]{kai09}
Kainulainen, J., Beuther, H., Henning, T., \& Plume, R. 2009, Astronomy and
  Astrophysics, 508, L35

\bibitem[{Kainulainen {et~al.}(2014)Kainulainen, Federrath, \& Henning}]{kai14}
Kainulainen, J., Federrath, C., \& Henning, T. 2014, Science, 344, 183

\bibitem[{Klessen {et~al.}(2000)Klessen, Heitsch, \& Mac~Low}]{kle00}
Klessen, R.~S., Heitsch, F., \& Mac~Low, M.-M. 2000, The Astrophysical Journal,
  535, 887

\bibitem[{{Launhardt} {et~al.}(2013){Launhardt}, {Stutz}, {Schmiedeke},
  {Henning}, {Krause}, {Balog}, {Beuther}, {Birkmann}, {Hennemann},
  {Kainulainen}, {Khanzadyan}, {Linz}, {Lippok}, {Nielbock}, {Pitann}, {Ragan},
  {Risacher}, {Schmalzl}, {Shirley}, {Stecklum}, {Steinacker}, \&
  {Tackenberg}}]{laun13}
{Launhardt}, R., {Stutz}, A.~M., {Schmiedeke}, A., {et~al.} 2013, \aap, 551,
  A98

\bibitem[{{Lombardi} {et~al.}(2015){Lombardi}, {Alves}, \& {Lada}}]{lombardi15}
{Lombardi}, M., {Alves}, J., \& {Lada}, C.~J. 2015, ArXiv e-prints

\bibitem[{{Lombardi} {et~al.}(2014){Lombardi}, {Bouy}, {Alves}, \&
  {Lada}}]{lombardi14}
{Lombardi}, M., {Bouy}, H., {Alves}, J., \& {Lada}, C.~J. 2014, \aap, 566, A45

\bibitem[{{Megeath} {et~al.}(2012){Megeath}, {Gutermuth}, {Muzerolle},
  {Kryukova}, {Flaherty}, {Hora}, {Allen}, {Hartmann}, {Myers}, {Pipher},
  {Stauffer}, {Young}, \& {Fazio}}]{megeath12}
{Megeath}, S.~T., {Gutermuth}, R., {Muzerolle}, J., {et~al.} 2012, \aj, 144,
  192

\bibitem[{Molina {et~al.}(2012)Molina, Glover, Federrath, \& Klessen}]{mol12}
Molina, F.~Z., Glover, S. C.~O., Federrath, C., \& Klessen, R.~S. 2012, Monthly
  Notices of the Royal Astronomical Society, 423, 2680

\bibitem[{{Myers} \& {Ladd}(1993)}]{myers93}
{Myers}, P.~C. \& {Ladd}, E.~F. 1993, \apjl, 413, L47

\bibitem[{{Nielbock} {et~al.}(2012){Nielbock}, {Launhardt}, {Steinacker},
  {Stutz}, {Balog}, {Beuther}, {Bouwman}, {Henning}, {Hily-Blant},
  {Kainulainen}, {Krause}, {Linz}, {Lippok}, {Ragan}, {Risacher}, \&
  {Schmiedeke}}]{nielbock12}
{Nielbock}, M., {Launhardt}, R., {Steinacker}, J., {et~al.} 2012, \aap, 547,
  A11

\bibitem[{{Ormel} {et~al.}(2011){Ormel}, {Min}, {Tielens}, {Dominik}, \&
  {Paszun}}]{ormel11}
{Ormel}, C.~W., {Min}, M., {Tielens}, A.~G.~G.~M., {Dominik}, C., \& {Paszun},
  D. 2011, \aap, 532, A43

\bibitem[{{Ossenkopf} \& {Henning}(1994)}]{ossen94}
{Ossenkopf}, V. \& {Henning}, T. 1994, \aap, 291, 943

\bibitem[{Padoan {et~al.}(2013)Padoan, Federrath, Chabrier, Evans, Johnstone,
  J{\o}rgensen, McKee, \& Nordlund}]{pad14}
Padoan, P., Federrath, C., Chabrier, G., {et~al.} 2013, arXiv.org

\bibitem[{Padoan {et~al.}(1997)Padoan, Nordlund, \& Jones}]{pad97}
Padoan, P., Nordlund, {\AA}., \& Jones, B. J.~T. 1997, Monthly Notices of the
  Royal Astronomical Society, 288, 145

\bibitem[{{Polychroni} {et~al.}(2013){Polychroni}, {Schisano}, {Elia}, {Roy},
  {Molinari}, {Martin}, {Andr{\'e}}, {Turrini}, {Rygl}, {Di Francesco},
  {Benedettini}, {Busquet}, {di Giorgio}, {Pestalozzi}, {Pezzuto},
  {Arzoumanian}, {Bontemps}, {Hennemann}, {Hill}, {K{\"o}nyves},
  {Men'shchikov}, {Motte}, {Nguyen-Luong}, {Peretto}, {Schneider}, \&
  {White}}]{polychroni13}
{Polychroni}, D., {Schisano}, E., {Elia}, D., {et~al.} 2013, \apjl, 777, L33

\bibitem[{{Roussel}(2013)}]{roussel13}
{Roussel}, H. 2013, \pasp, 125, 1126

\bibitem[{{Sadavoy}(2013)}]{sad13}
{Sadavoy}, S.~I. 2013, PhD thesis, University of Victoria

\bibitem[{{Sadavoy} {et~al.}(2014){Sadavoy}, {Di Francesco}, {Andr{\'e}},
  {Pezzuto}, {Bernard}, {Maury}, {Men'shchikov}, {Motte},
  {Nguy{\tilde}{\^e}n-Lu'o'ng}, {Schneider}, {Arzoumanian}, {Benedettini},
  {Bontemps}, {Elia}, {Hennemann}, {Hill}, {K{\"o}nyves}, {Louvet}, {Peretto},
  {Roy}, \& {White}}]{sadavoy14}
{Sadavoy}, S.~I., {Di Francesco}, J., {Andr{\'e}}, P., {et~al.} 2014, \apjl,
  787, L18

\bibitem[{{Safron} {et~al.}(2015){Safron}, {Fischer}, {Megeath}, {Furlan},
  {Stutz}, {Stanke}, {Billot}, {Rebull}, {Tobin}, {Ali}, {Allen}, {Booker},
  {Watson}, \& {Wilson}}]{safron15}
{Safron}, E.~J., {Fischer}, W.~J., {Megeath}, S.~T., {et~al.} 2015, \apjl, 800,
  L5

\bibitem[{Schneider {et~al.}(2013)Schneider, Andr{\'e}, K{\"o}nyves, Bontemps,
  Motte, Federrath, Ward-Thompson, Arzoumanian, Benedettini, Bressert, Didelon,
  Di~Francesco, Griffin, Hennemann, Hill, Palmeirim, Pezzuto, Peretto, Roy,
  Rygl, Spinoglio, \& White}]{sch13}
Schneider, N., Andr{\'e}, P., K{\"o}nyves, V., {et~al.} 2013, The Astrophysical
  Journal, 766, L17

\bibitem[{{Sodroski} {et~al.}(1997){Sodroski}, {Odegard}, {Arendt}, {Dwek},
  {Weiland}, {Hauser}, \& {Kelsall}}]{sodroski97}
{Sodroski}, T.~J., {Odegard}, N., {Arendt}, R.~G., {et~al.} 1997, \apj, 480,
  173

\bibitem[{{Stanke} {et~al.}(2010){Stanke}, {Stutz}, {Tobin}, {Ali}, {Megeath},
  {Krause}, {Linz}, {Allen}, {Bergin}, {Calvet}, {di Francesco}, {Fischer},
  {Furlan}, {Hartmann}, {Henning}, {Manoj}, {Maret}, {Muzerolle}, {Myers},
  {Neufeld}, {Osorio}, {Pontoppidan}, {Poteet}, {Watson}, \&
  {Wilson}}]{stanke10}
{Stanke}, T., {Stutz}, A.~M., {Tobin}, J.~J., {et~al.} 2010, \aap, 518, L94

\bibitem[{{Stutz} {et~al.}(2010){Stutz}, {Launhardt}, {Linz}, {Krause},
  {Henning}, {Kainulainen}, {Nielbock}, {Steinacker}, \& {Andr{\'e}}}]{stutz10}
{Stutz}, A., {Launhardt}, R., {Linz}, H., {et~al.} 2010, \aap, 518, L87

\bibitem[{{Stutz} {et~al.}(2013){Stutz}, {Tobin}, {Stanke}, {Megeath},
  {Fischer}, {Robitaille}, {Henning}, {Ali}, {di Francesco}, {Furlan},
  {Hartmann}, {Osorio}, {Wilson}, {Allen}, {Krause}, \& {Manoj}}]{stutz13}
{Stutz}, A.~M., {Tobin}, J.~J., {Stanke}, T., {et~al.} 2013, \apj, 767, 36

\bibitem[{{van Kempen} {et~al.}(2009){van Kempen}, {van Dishoeck}, {Salter},
  {Hogerheijde}, {J{\o}rgensen}, \& {Boogert}}]{vankempen09}
{van Kempen}, T.~A., {van Dishoeck}, E.~F., {Salter}, D.~M., {et~al.} 2009,
  \aap, 498, 167

\end{thebibliography}

\appendix

\section{Column density map of Orion A} \label{sec:ap_nh}

We have generated reduced data products for the Orion A region using
HIPE processing to level~1 followed by final level~2 Scanamorphos
processing \citep[version 24.0, using the ``galactic'' option,
][]{roussel13}.  The column density (and temperature) maps were
derived in a similar way as those presented in
\citet{stutz10,stutz13}, and \citet{laun13}, with the total power
emission levels at each of the four wavelengths calculated using
Planck and IRAS data \citep{bernard10}.  We briefly summarize the
steps we use to generate the N(H) and temperature maps here and refer
the reader the above works for more details.  We apply the total power
emission level corrections for each of the 160~\micron, 250~\micron,
350~\micron, and 500~\micron intensity maps \citep{bernard10}. We
convolve the data to the largest beam: the 500~\micron beam, with
$\sim\,$38\arcsec\ FWHM.  We use the azimuthally averaged convolution
kernels from \citet{aniano11}. We then re--grid the data to a matched
coordinate system and pixel scale, in this case we adopt an 18\arcsec
pixel scale (see Appendix~\ref{sec:ap_fit} for more details). The conversion
to matched units assumes the beam sizes listed in the SPIRE instrument
handbook.  We fit the spectral energy distribution (SED) of each pixel
assuming a modified black--body function of the form
\begin{equation}
  S_{\nu} = \Omega\,B_{\nu}(\nu,T_{\rm d})\,(1-e^{-\tau(\nu)}),
\end{equation}
where $\Omega$\ is the solid angle of the emitting element,
$B_{\nu}(T_{\rm d})$ is the Planck function at a dust temperature
$T_{\rm d}$, and $\tau(\nu)$\ is the optical depth at frequency $\nu$.
Here, the optical depth is given by $\tau(\nu) = N_{\rm H}\,m_{\rm
  H}\,R_{gd}^{-1}\,\kappa(\nu)$, where $N_{\rm H} = 2\,\times\,N({\rm
  H_2}) + N({\rm H})$ is the total hydrogen column density, $m_{\rm
  H}$ in the proton mass, $\kappa_{\nu}$ is the assumed frequency
dependent dust opacity law, and $R_{gd}$ is the gas--to--dust ratio,
assumed to be 110~\citep{sodroski97}.  We use the \citet{ossen94}
model dust opacities corresponding to column 5 of their Table~1
(sometimes referred to as the ``OH5'' opacities).  These opacities are
meant to reflect grains with thin ice mantles after $10^5$~years of
coagulation time at an assumed gas density of $10^6$~cm$^{-3}$.  See
\citet{stutz13} and \citet{laun13} for discussions on the systematic
uncertainties introduced by the model dust opacity assumption.  In
order to apply the color and beam size corrections recommended in the
SPIRE and PACS instrument handbooks, we first fit the uncorrected
pixel SED to estimate the temperature.  We then apply the interpolated
correction value at that temperature and re--fit the SED.

\begin{figure*}
  \centering
  \scalebox{0.8}{\includegraphics[trim = 14mm 0mm 0mm 0mm, clip]{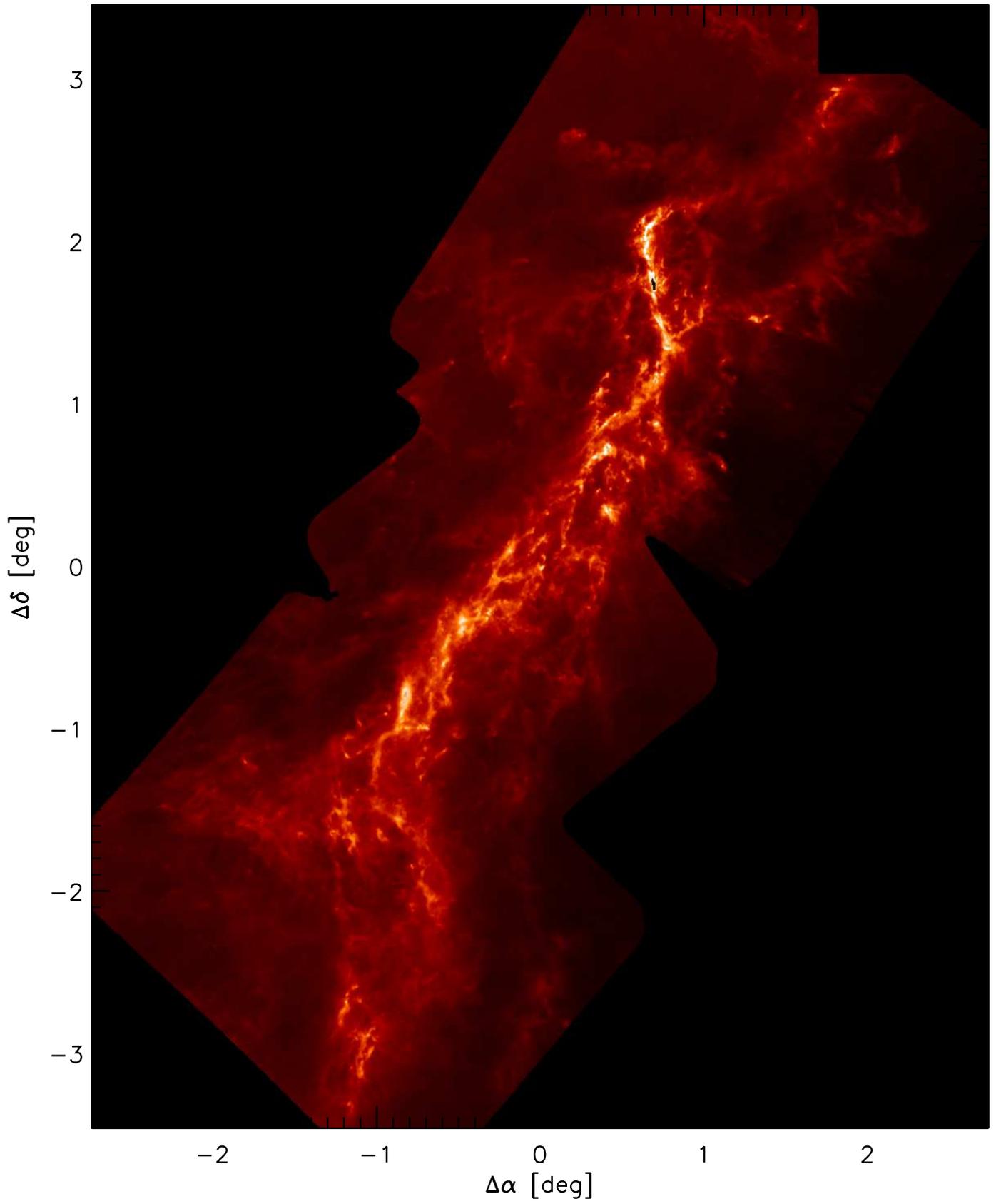}}
  \caption{Orion~A N(H) map, shown on a log scale.}
  \label{fig:ap_nh}
\end{figure*}

In an additional step, we use the 500~\micron resolution temperature
($T_{\rm d}$) map to convert the 250~\micron intensity map to a column
density map in order to improve the final resolution
(Figs.~\ref{fig:ap_nh} and \ref{fig:nh}).  Both
maps compare well; only minor differences are apparent on the smallest
scales caused by resolution effects, as expected.  The previously
published \herschel--derived N(H) maps compare well to the maps we
present here \citep[e.g.,][]{lombardi14,polychroni13}.

\section{Fitting N(H) PDFs without binning} \label{sec:ap_fit}

Here we describe the method we use for fitting an index $\alpha$ to a
power--law distribution of the form $dN/dZ \propto Z^\alpha$ over a
finite range in $Z$ values from $Z_\min$ to $Z_\max$.
The differential probability of column density $Z_i$,
given a power--law distribution $Z^\alpha$ and total number of expected
detections $n_{exp}$ in the interval between $Z_\min$ and $Z_\max$ is
$$
P_i = Z_i^\alpha {x\over Z_\max^x - Z_\min^x} n_\ex.
\qquad (x \equiv \alpha + 1)
$$
The likelihood ${\cal L}$ is given by the product of the individual
probabilities, or
$$
\ln {\cal L} = \sum_{i=1}^{n_\obs} \ln P_i = 
\alpha \sum_{i=1}^{n_\obs} \ln Z_i + n_\obs(\ln x - \ln(Z_\max^x - Z_\min^x)
+ \ln n_\ex). 
$$
It is straightforward to show that the likelihood is maximized when
the models are restricted to those with $n_\ex=n_\obs$.  Hence, we
simplify notation by $n=n_\ex=n_\obs$ and obtain
$$
\ln {\cal L} =  
\alpha \sum_{i=1}^{n} \ln Z_i + n(\ln x - \ln(Z_\max^x - Z_\min^x) 
+ \ln n)
$$
To find where ${\cal L}$ is maximized, we differentiate w.r.t.\ $\alpha$
(or $x=\alpha+1$) and set to zero
$$
0 = \sum_{i=1}^n \ln Z_i + {n\over x} 
- n{Z_\max^x\ln Z_\max - Z_\min^x\ln Z_\min \over Z_\max^x - Z_\min^x},
$$
which can be rewritten as
$$
\langle \ln Z\rangle =  \ln Z_\min {r R^x - 1\over R^x -1} - {1\over x};
$$
where
$$
\langle \ln Z\rangle \equiv {1\over n}\sum_{i=1}^n \ln Z_i,\quad
r \equiv {\ln Z_\max\over\ln Z_\min},\quad
R \equiv {Z_\max\over Z_\min}.
$$
The error is given by \citet{gould95} Eq.\ 2.4: 
$$
{1\over \sigma(x)} = \sqrt{n\biggl[{1\over x^2} - {R^{-x}(\ln R)^2\over 
(R^{-x} - 1)^2}\biggr]}
$$
which can be written more elegantly as
$$
{1\over \sigma(x)} = {1\over |x|}
\sqrt{n\biggl[1 - \biggl({Q\over \sinh Q}\biggr)^2\biggr]},
\qquad
\biggl(Q\equiv {x\ln R\over 2}\biggr)
$$
This expression can be Taylor expanded:
\begin{equation}
\sigma(x) = {1\over \ln R}\sqrt{12\over n}\biggl(1 + {(x\ln R)^2\over 40}
+ \dots \biggr).
\label{eqn:erra}
\end{equation}

Equation~\ref{eqn:erra} represents the analytical error solution
assuming the power-law model accurately reflects the distribution of
data values.  Therefore it represents a lower limit for the errors.
Alternatively, the errors can be estimated using the $\Delta\,\chi^2 =
1.0$ interval, which compares well to those derived using the minimum
variance bound (see~Table~\ref{tab:ap_err} for a comparison between
the two).  The power--law indices and $\Delta\,\chi^2 = 1.0$ errors
are listed in Table~\ref{tab:sf}.  

\begin{table}
\caption{Power--law index error comparison}
\begin{center}
\begin{tabular}{cccc}
  \hline \hline
  Region &
  Index &
  $\Delta\,\chi^2$ &
  $\sigma\,^a$ \\
  &
  &
  error & 
   \\
  \hline
  1 & -0.93 & 0.039 & 0.042 \\
  2 & -1.03 & 0.036 & 0.035 \\
  3 & -1.95 & 0.047 & 0.038 \\
  4 & -2.26 & 0.062 & 0.053 \\
  5 & -2.04 & 0.061 & 0.044 \\
  6 & -2.30 & 0.090 & 0.067 \\
  7 & -1.91 & 0.047 & 0.038 \\ 
  8 & -2.95 & 0.080 & 0.067 \\
  \hline
\end{tabular}
\end{center}
\footnotesize{The errors reported here correspond to those derived
  from the N(H) intervals listed in Table~\ref{tab:sf}. $^a$ Error
  from equation~\ref{eqn:erra}}
\label{tab:ap_err}
\end{table}

\subsection{The effects of resolution, pixel--size, and fitting range on the power--law index}

\begin{figure}
  \centering
  \scalebox{0.5}{\includegraphics[trim = 0mm 0mm 0mm 0mm, clip]{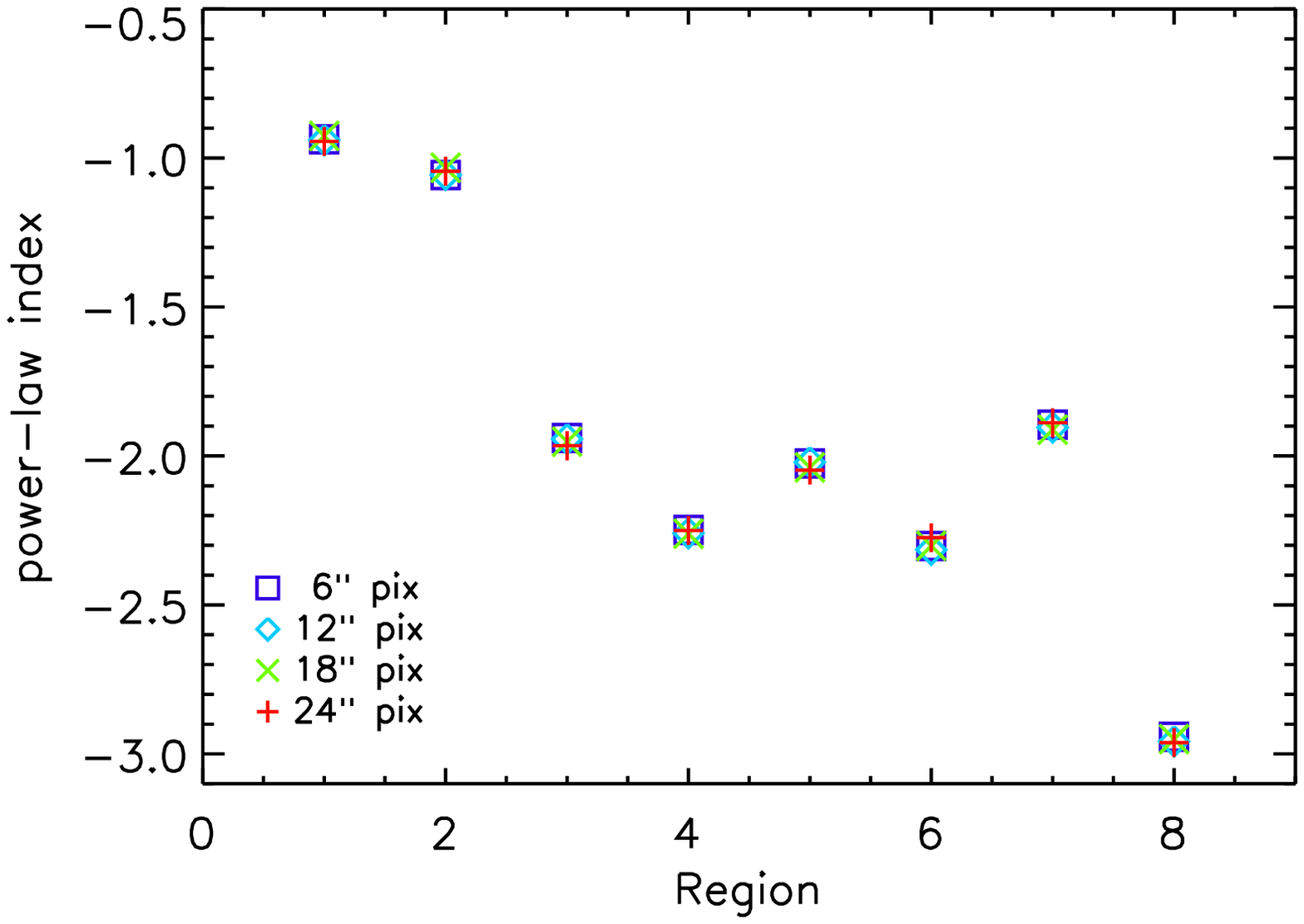}}
  \scalebox{0.5}{\includegraphics[trim = 0mm 0mm 0mm 10mm, clip]{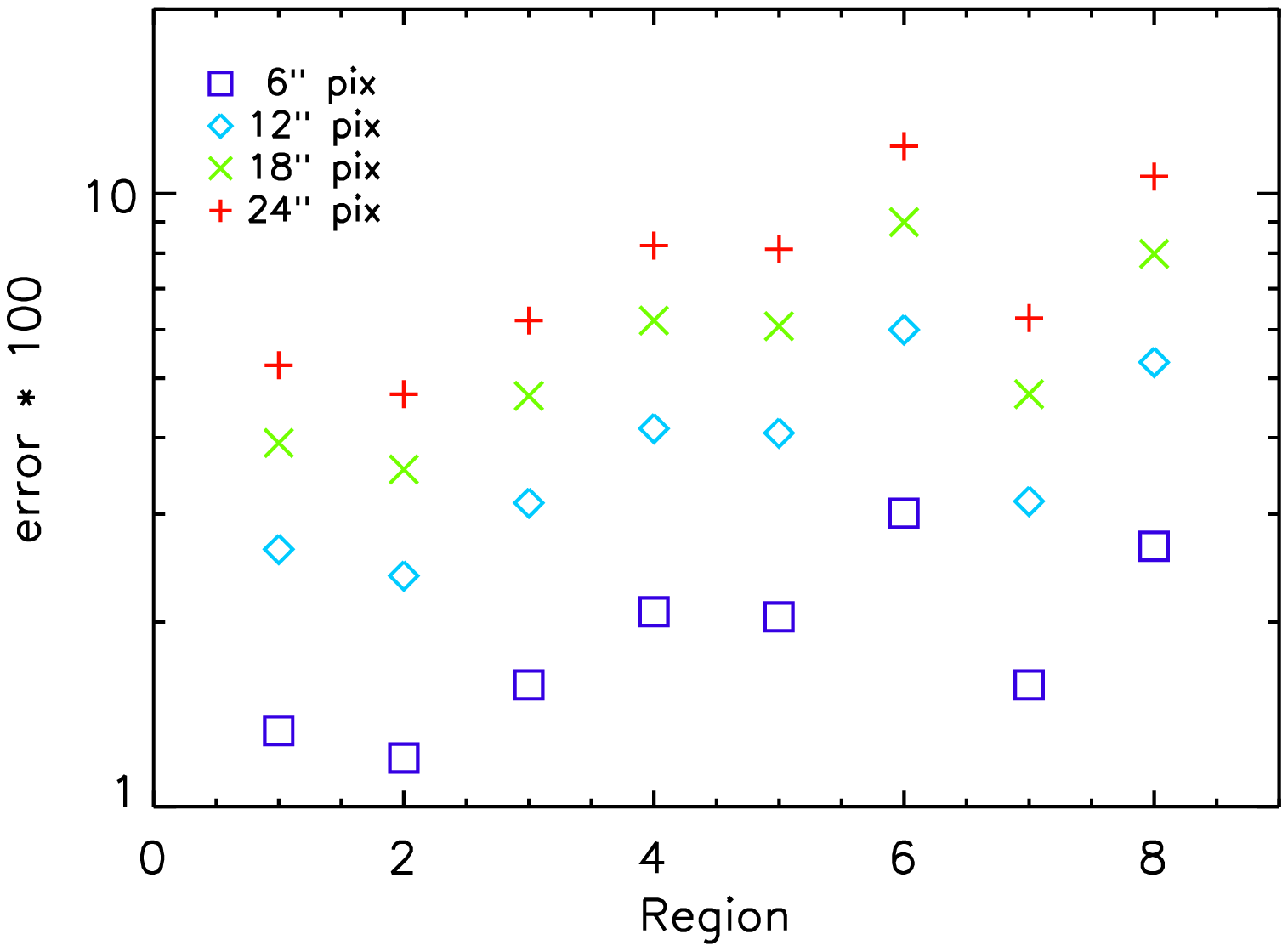}}
  \caption{Top: Best-fit power-law index for each region as a function
    of N(H) map pixel size.  The slope values are not affected by
    correlated pixel, and differences are 2\% at most. Bottom: Error
    for each region as a function of N(H) map pixel size.  The error
    increases with pixel size because number of counts decreases.  All
    quantities shown here have been derived using the same N(H) limits
    presented in Table~\ref{tab:sf}.}
  \label{fig:ap_250slopetest}
\end{figure}

\begin{figure}
  \centering
  \scalebox{0.5}{\includegraphics[trim = 0mm 0mm 0mm 0mm, clip]{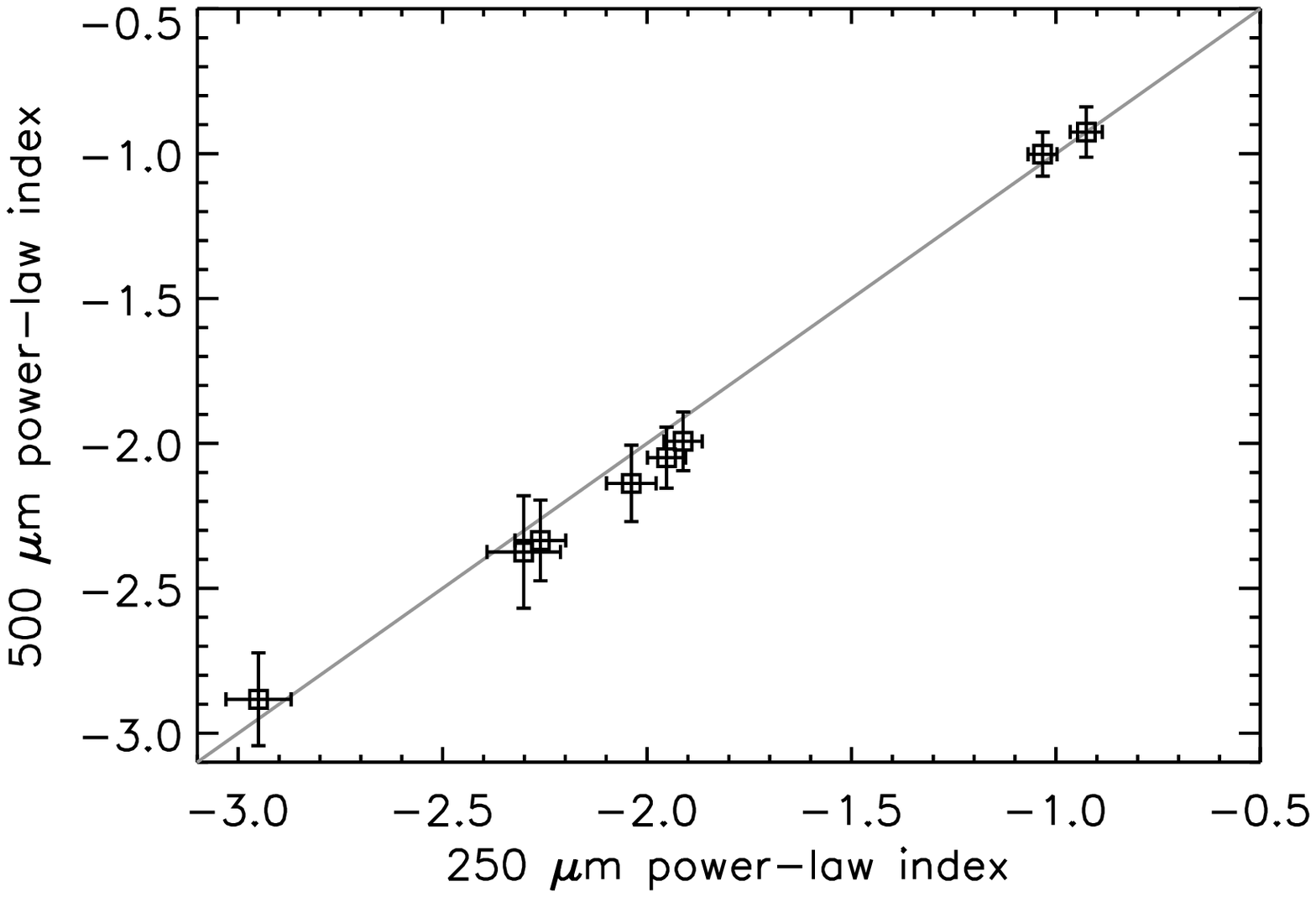}}
  \caption{Indices derived from the 250~\micron 18\arcsec pixel map
    vs.\ those derived from the 500~\micron 40\arcsec pixel map, both
    using the N(H) limits presented in Table~\ref{tab:sf}.}
  \label{fig:ap_slopecomp}
\end{figure}

As described above, the power--law indices for each region are
extracted from 18\arcsec\ pixel N(H) maps derived from the 250~\micron
map of Orion~A.  Here we test the effect of adopting different pixel
scales for the 250~\micron N(H) map.  As shown in
Fig.~\ref{fig:ap_250slopetest} the adopted pixel scale has a
negligible effect on the indices, with a maximum effect on the
best-fit indices of $\sim2$\%.  The fractional errors increase with
pixel size because the number of pixels decreases.  We find similar
results using the 500~\micron N(H) map (with a beam size of
$\sim\,$36\arcsec\ FWHM) and adopting pixel sizes of 10\arcsec,
20\arcsec, 30\arcsec, and 40\arcsec: the power--law index is only
marginally affected by the choice in pixel size. The fractional errors
exhibit the same behavior as for the 250~\micron, but are somewhat
larger (maximum of 10\% at 40\arcsec) due to the smaller number of
pixels available to fit. Finally, in Fig.~\ref{fig:ap_slopecomp} we
compare the indices derived from the 250~\micron 18\arcsec\ pixel map
with those derive from the 500~\micron 40\arcsec\ pixel map.  We find
good agreement between the two slope estimates.

We also test varying the upper bound of the fitting range shown as Max
log N(H) in Table~\ref{tab:sf}. For a change $\pm \Delta \log \rm{N(H)} =
0.3$ we find the indices change by $\sim$\,10\% or less, a variation
that is much smaller than differences between regions.  

\section{The effects of completeness and extinction on the Class~0 protostar fraction}\label{sec:ap_comp}
\subsection{Completeness of protostellar catalogs}

\begin{figure}
  \centering
  \scalebox{0.5}{\includegraphics[trim = 0mm 0mm 0mm 0mm, clip]{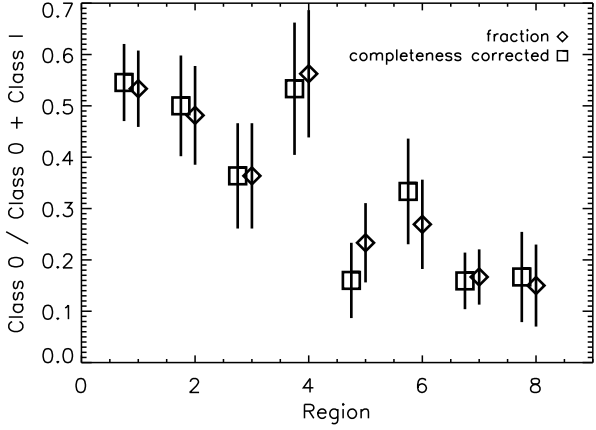}}
  \caption{Class~0 fraction in regions.  Diamonds indicate the total
    fraction (Table~\ref{tab:sf}), while squares indicate the
    completeness corrected values.  Results remain consistent with and
    without the application of the completeness limit.}
  \label{fig:ap_fraccomp}
\end{figure}

The northern portion of Orion~A (regions 1 and 2) are those most
affected by incompleteness due to elevated levels of nebulosity.  We
estimate the completeness limit in the HOPS protostellar catalog as
follows. We inject artificial sources into the 70~\micron\ images and
recover them with our source--finding software \citep[see ][for
  details]{stutz13}.  We estimate the mean 90\% completeness level for
regions 1 and 2 to be 0.12~Jy, while for the rest of L1641 it is
0.03~Jy. We therefore apply a 0.12~Jy 70~\micron\ flux limit to the
HOPS protostar catalog, eliminating about 10\% of protostars from the
catalog. We then calculate Class~0 fractions from the remaining
sources.  We find excellent agreement between the completeness
corrected and total raw sample fractions, as shown in
Fig.~\ref{fig:ap_fraccomp}.

\subsection{The effect of foreground extinction on \tbol}

Our goal is to assess the effects of foreground extinction on the
\tbol--based YSO classification.  We assume here that the
main contribution to an erroneous \tbol classification is the
misidentification of Class~I sources as Class~0 protostars (\tbol)
$< 70$~K).  While other sources of contamination are also possible,
scrutiny of the HOPS protostellar SEDs reveals that the current
classification based on observed colors and fluxes, in combination
with PACS and 870~\micron data, is robust (Furlan et al., in
preparation).

\begin{table}
\caption{Upper limits to the distribution of protostellar extinction per region}
\begin{center}
\begin{tabular}{cccc}
  \hline \hline
  Region &
  Median$^a$ &
  Minimum$^a$ &
  Maximum$^a$ \\
  &
  A$_{\rm V}$ &
  A$_{\rm V}$ & 
  A$_{\rm V}$ \\
  \hline
  1 & 23.3 & 0.64 & 35.6 \\
  2 & 13.3 & 2.18 & 30.0 \\
  3 & 16.8 & 4.34 & 31.8 \\ 
  4 & 7.51 & 4.08 & 26.4 \\
  5 & 13.0 & 2.63 & 21.2 \\ 
  6 & 14.9 & 4.03 & 29.7 \\
  7 & 12.5 & 3.65 & 35.1 \\
  8 & 8.79 & 4.54 & 16.2\\ 
\hline
\end{tabular}
\end{center}
\footnotesize{$^a$ Median, minimum, and maximum A$_{\rm V}$ (mag) from the
  calculated from the \herschel N(H) maps centered on the positions of
  protostars in each region. A$_{\rm V}$ are calculated in an annulus
  of size 18\arcsec $\times [1.5, 4 pixels]$ (or $\sim$ 11400~AU to
  30300~AU).}
\label{tab:ap_ext}
\end{table}

\begin{figure}
  \centering
  \scalebox{0.53}{\includegraphics{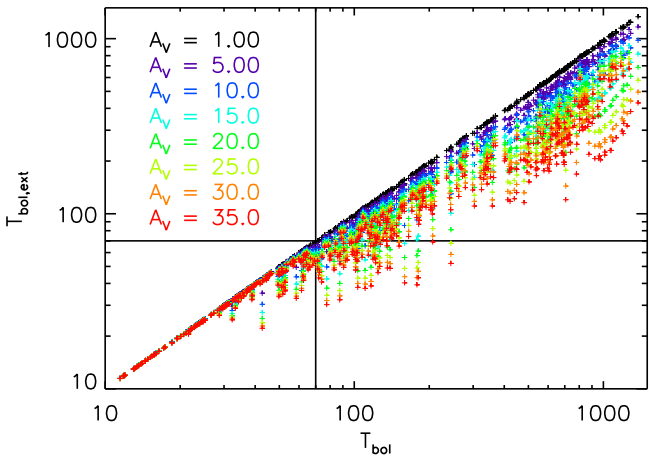}}
  \caption{Model \tbol values versus SED extincted \tbole values.
    Colors indicate the assumed levels of extinction.  Here we show
    500 randomly chosen models. As the extinction A$_{\rm V}$
    increases, the model \tbole decreases and the levels of
    contamination at \tbole$<$ 70 K increases.}
  \label{fig:ap_tbolv}
\end{figure}

We obtain observational constraints on the levels of foreground
extinction toward individual protostars from the \herschel N(H) map.
We measure the N(H) toward each protostar using an annulus of size
18\arcsec $\times [1.5,4]$ (or $\sim$ 11400~AU to 30300~AU), adopted
to avoid extinction intrinsic to the protostellar envelopes.  Within
each annulus we calculate the median value of N(H). In
Table~\ref{tab:ap_ext} we list the median, minimum, and maximum
protostellar N(H) values for each region, assuming a conversion of
A$_{\rm V}$/N(H)$= 1.85 \times 10^{21} {\rm mag\,cm}^{-2}$ \citep[e.g., ][,
derived for the diffuse ISM, and used here for notational
simplicity]{bohlin78}. These values are upper limits because the N(H)
map integrates the extinction through the entire cloud.  The
protostellar A$_{\rm V}$ values have a maximum of 35 mag in regions 1
and 7.  However, more typical values range from $\sim$20 mag to
$\sim$10 mag.

\begin{figure}
  \centering
  \scalebox{0.53}{\includegraphics[trim = 0mm 0mm 0mm 0mm, clip]{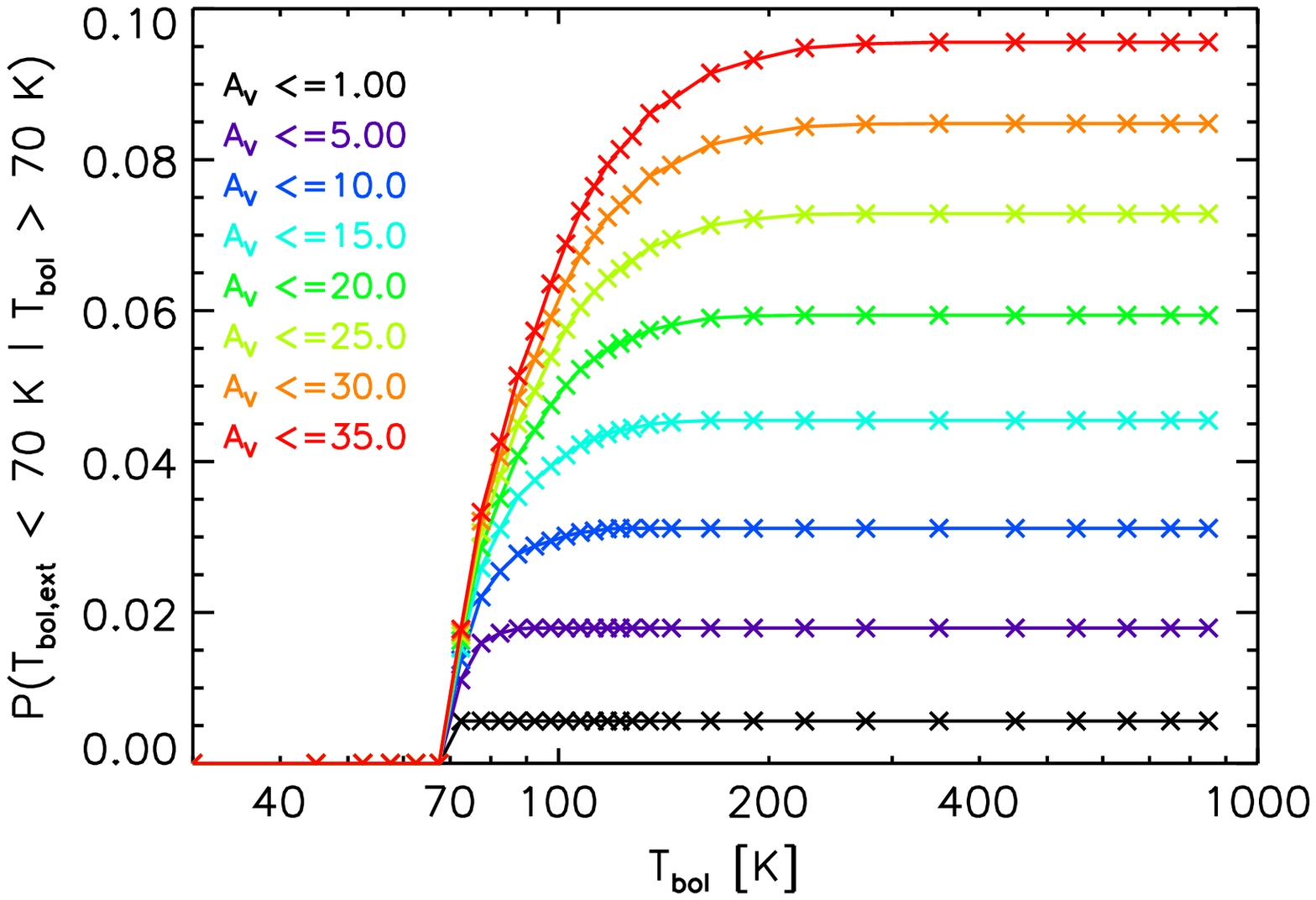}}
  \scalebox{0.53}{\includegraphics[trim = 0mm 0mm 0mm 10mm, clip]{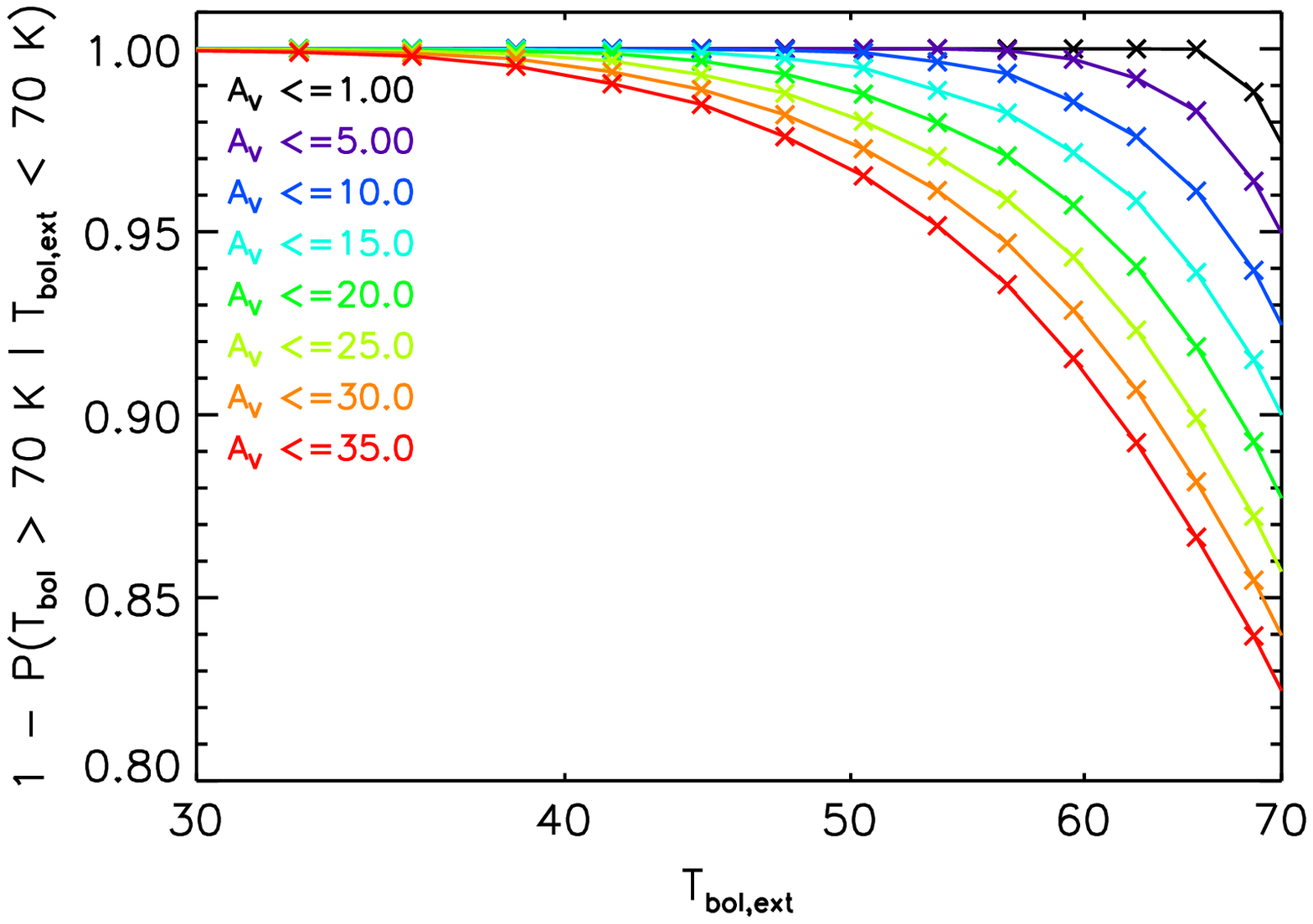}}
  \caption{Top: Probability that a source with a given \tbol $>$70~K
    will have \tbole$<$70~K, marginalizing over a uniform foreground
    A$_{\rm V}$ extinction distribution up to the listed value (shown
    in color).  There is no Class~0 contamination from Class~I models
    with \tbol$>$ 400~K for the A$_{\rm V}$ values we consider here.
    Bottom: Probability that a source is correctly classified as a
    Class~0 source (has \tbol $<$ 70~K).}
  \label{fig:ap_tbol}
\end{figure}

In order to assess the levels of contamination in the Class 0 sample
we use the protostellar model grid presented in \citet{stutz13}. We
refer the reader to that publication and to Furlan et al.\ (in prep.)
for more details. In brief, we vary 5 parameters in our grid:
inclination of the disk relative to the LOS, the envelope density, the
cavity opening angle, the disk size, and the luminosity of the central
protostar. Our grid contains uniformly sampled paramaters and uses the
\citet{ormel11} model dust opacities (``icsgra3''). In total, our grid
contains 30400 individual models.  About $\sim$30\% of the models have
\tbol$<$70~K, similar to the observed protostellar distribution.  We
attenuate each model SED (adopting the wavelength coverage of the
observations) with a range of A$_{\rm V}$ values between 1 and 35~mag,
spaning the observed range in A$_{\rm V}$ toward the protostar sample
(Table~\ref{tab:ap_ext}).  For consistency with the above protostellar
YSO measurements, we attenuate the model SEDs with the same model as
that used to derive the N(H) map \citep[][; ``OH5'']{ossen94},
assuming A$_{\rm V}$/A$_{\rm K} = 14$ \citep{nielbock12}.  In
Fig.~\ref{fig:ap_tbolv} we show resulting extincted \tbole values as a
function of \tbol for a random subset of models.  We then measure the
fraction of models that have \tbol~$>$~70~K and \tbole$<$ 70~K as a
function of \tbol and A$_{\rm V}$, marginalizing over a uniform
A$_{\rm V}$ distribution.  The probability that a given Class~I
protostar is masquerading as Class~0 protostar is shown in
Fig.~\ref{fig:ap_tbol}, top panel.  The bottom panel of
Fig.~\ref{fig:ap_tbol} shows the probability that a given Class~0
protostar is correctly identified (has \tbol$<$70~K).

We obtain a maximum fraction of $\sim$\,9\% of Class~I sources that
may appear as Class~0 protostars if models are extincted by up to
A$_{\rm V}$ = 35 mag.  For all models with \tbol$>$400~K, there is
zero contamination of Class~0 sources up to extinctions of
A$_{\rm V}\sim60$~mag. We conclude that the main source of
extinction--driven contamination on the Class~0 sample is therefore
from Class I sources with 70 K$<$\tbol$<$400~K.  We also find that
there is $<5\%$ contamination of Class~0 protostars below
\tbole$\sim$45~K and below foreground extinction levels of
A$_{\rm V}$~=~35~mag. For a median extinction value of
A$_{\rm V}$~=~20~mag, the fraction of Class~0 contamination is less
than 5\% for \tbole$<$55~K.

\begin{figure}
  \centering
  \scalebox{0.53}{\includegraphics{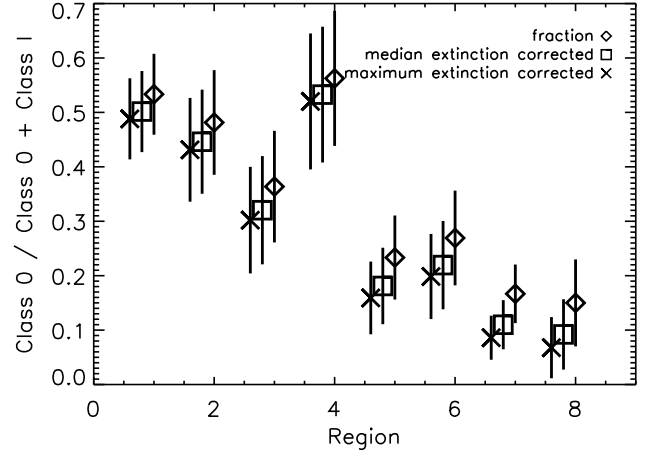}}
  \caption{Dependence of the fraction of Class~0 protostars on \tbol
    misclassification due to foreground extinction.  Diamonds indicate
    the raw uncorrected observed fractions. Squares indicate the
    fractions derived by correcting for the sample median A$_{\rm V}$
    extinction levels toward protostars in each region.  X-symbols
    indicate the fractions derived by correcting for the maximum
    A$_{\rm V}$ observed toward protostars in each region.}
  \label{fig:ap_cfrac}
\end{figure}

Using the total contamination fractions shown in
Fig.~\ref{fig:ap_tbol} and both the median and maximum observed
A$_{\rm V}$ levels (Table~\ref{tab:ap_ext}), we apply correction
factors to the observed numbers of Class~0 protostars, keeping the
total number of YSOs fixed at the observed value.  The dependence of
the Class~0 fraction on extinction contamination is presented in
Fig.~\ref{fig:ap_cfrac}.  While the overall amplitude of the fraction
is affected by the foreground extinction effects, the shape is not.
As noted above, the bottom panel of Fig.~\ref{fig:ap_tbol}
demonstrates that at \tbole$<$55~K the amount of Class~0 contamination
is less than 5\% for our expected levels of extinction. We therefore
test the effects of adopting a 55~K division between Class~0 and
Class~I protostars.  We find that the results are similar to those
shown in Fig.~\ref{fig:ap_cfrac} for the median extinction corrected
values, as expected.  We therefore conclude that contamination driven
by foreground extinction cannot account for the observed trend
presented in Fig.~\ref{fig:cor}.

We note that our requirement of a PACS 70~\micron detection and the
inclusion of longer wavelengths covering the peak of the cold--dust
SED, in combination with meticulous short wavelength selection
\citep{megeath12}, virtually guarantees the presence of an envelope.
Therefore we do not consider Class~I protostar contamination from more
evolved sources to be a significant source or error \citep[see also
][]{heiderman15}.

\end{document}